\documentclass[11pt]{article} 
\usepackage[textheight=216mm,textwidth=160mm]{geometry}
\usepackage{graphicx} 
\usepackage{natbib}
\usepackage{enumerate}
\usepackage{caption}

\usepackage{multirow}
\usepackage{makecell}
\usepackage{hyperref}
\hypersetup{
    colorlinks=true,
    linkcolor=blue,
    filecolor=blue,      
    urlcolor=blue,
    citecolor=black
    }

\usepackage{newtxtext}
\usepackage{newtxmath}

\title{Surfactant effect on collective bubble bursting and aerosol emission}

\author{Megan Mazzatenta$^1$, Samuel M. Koblensky $^1$, Luc Deike$^{1,2}$\thanks{email for correspondance: ldeike@princeton.edu}}

\date{$^1$ Department of Mechanical and Aerospace Engineering, Princeton University, Princeton, NJ, 08544, USA\\
$^2$High Meadows Environmental Institute, Princeton University, Princeton, NJ, 08544, USA}

\begin{document}
\maketitle

\begin{abstract}

Bubbles entrained by breaking waves rise to the ocean surface where they cluster and burst, emitting sea spray aerosols into the atmosphere. Bubble bursting thereby links seawater biogeochemistry and aerosol chemistry, influencing the ability of emitted aerosols to serve as cloud condensation nuclei or ice nucleating particles. The mechanisms of film drop and jet drop production are modulated by organic material present in seawater, which may affect the size, number, and composition of resulting aerosols. 
We disentangle the effect of surfactant on collective bursting processes using laboratory experiments with detailed bubble and aerosol measurements down to small sizes, multiple bubble size configurations, and measurements of bubble lifetime. 
Submicron aerosol emission, linked to film drop production, increased with surfactant up to an optimal concentration, while production of supermicron aerosols emitted through jet drop production was shut down.
Our work paves the way to integrate organic composition into sea spray emission functions.

\end{abstract}

\section{Introduction}

Bubble bursting at the ocean surface couples ocean biogeochemistry and atmospheric chemistry by emitting sea spray aerosols composed of salts and organic material. The generated aerosols significantly influence atmospheric processes, especially in remote marine regions, by modulating the radiative balance of the atmosphere and serving as cloud condensation nuclei or ice nucleating particles \citep{gantt_rev,burrows_oceanfilms_2014,painemal_2015,cochran_sea_2017,kirpes}. The ability of aerosols to nucleate drops and ice particles is highly dependent on their size as well as on their chemical composition \citep{fuentes2,de_leeuw_production_2011,quinn_chemistry_2015,demott_inp,wang2017,xu_ccn,burrows_oceanfilms_2022}, which both depend on the specific mode of drop production by bursting bubbles, namely film drop or jet drop production \citep{prather2013,collins,wang2017}.

Accurately characterizing aerosol emissions in atmospheric and climate models is therefore critical, especially for representing cloud formation in the Arctic and Southern Oceans. However, significant uncertainties remain in the parameterization of sea spray emission functions (reviewed in \citet{lewis_sea_2004, de_leeuw_production_2011,veron_ocean_2015}), which has motivated the development of physics-based formulations to describe the sea spray aerosol size distribution, building off of laboratory experiments and associated fundamental understanding \citep{deike_mechanistic_2022}. Other work has investigated the link between seawater composition and aerosol composition, both theoretically \citep{burrows_oceanfilms_2014,burrows_2016, burrows_oceanfilms_2022} and experimentally \citep{fuentes,prather2013,kirpes}. Due to the complex chemistry of the ocean and challenges in making detailed open ocean measurements of bubbles and aerosols, controlled laboratory-scale experiments are needed to understand the physics of aerosol emission in organic solutions, and eventually connect to the aerosol composition. 

Quantifying the effect of organics on bubble bursting and aerosol emission has remained an open challenge, as surfactants influence several bubble processes through their ability to reduce the static surface tension at the air-water interface, induce Marangoni flows by creating gradients in surface tension along the bubble cap or cavity, interact with the bulk solution through dynamic adsorption-desorption processes, and interact with ions in salt solutions (as reviewed in \citet{manikantan_review}, see additionally \citet{Fernandez_adsorption} for adsorption time effects and summary in \citet{salt_ion_summary} for ion interactions). Regarding the effect on specific bubble processes, surfactants modify the size distribution of bubbles created underwater by breakup in turbulence \citep{zhan_tristan}, affect bubble rise velocity (\citet{wang_surfact_risevel,farsoiya_coupled_2024}, and review by \citet{surfact_risevel}), extend the lifetime of surface bubbles \citep{modini_effect_2013,poulain_ageing_2018,tristan_bubble}, somehow modulate jet and film drop production mechanisms \citep{constante,pierre, pico, vega, jun_tristan}, and lead to enrichment of organic material in aerosols \citep{odowd_biogenic,fuentes2,prather2013,wang2017,kirpes,kimble_soars}. When looking at collective bubbles (bubbles in clusters at the surface), surfactants prevent coalescence and increase the lifetime and size of bubble rafts, which can lead to the formation of tightly-packed rafts of multilayer foams in highly-contaminated conditions \citep{garrett_stabilize, weaire1999physics,cantat2013foams}.

Previous experiments have measured the aerosols produced by collective bubbles in a variety of organic solutions, both in studies with laboratory surfactants and using natural seawater samples \citep{cipriano_bubble_1981,martensson_laboratory_2003,l-b1_2003,sellegri_surfactants_2006,tyree_foam_2007,modini_effect_2013,prather2013, wang2017, frossard_marine_2019}. Many of these studies observed a shift to smaller mean aerosol sizes (focusing primarily on submicron aerosols) for various organic solutions as contamination was increased (including \citet{sellegri_surfactants_2006} with sodium dodecyl sulfate (SDS) solutions, \citet{modini_effect_2013} using Triton-X, \citet{tyree_foam_2007} using oleic acid, and \citet{fuentes} using filtered seawater solutions). However, the studies found conflicting results regarding the number of emitted aerosols, with most observing some amount of increase in production of submicron aerosols with surfactant. Without both detailed bubble measurements and precise knowledge of the composition of the solution, it remains difficult to compare results from separate studies and to distinguish the effect of surfactant on the physical bursting process from differences due to varied bubble size distributions or collective bursting.

Adding on detailed bubble measurements, works from \citet{neel2021}, \citet{neel_surf_2022}, and \citet{mazzatenta} present a bubbling tank setup designed for systematic measurement of underwater bubbles, surface bubbles, and drops in controlled solutions to directly link collective bursting bubbles to the aerosols they emit. \citet{neel2021,neel_surf_2022} considered bubbles in deionized water with surfactant and measured drops above 30~\textmu{m}, finding that an intermediate surfactant concentration led to an optimal production efficiency of jet drops, which then decreased once the surface became packed with bubbles. 
In the experiments of \citet{mazzatenta}, we extend the measurements of both bubbles and drops to much smaller sizes to make a quantitative link between bursting bubbles and emitted drops. We show that for inorganic artificial seawater solutions, the size and number of drops emitted by collective bursting bubbles can be described through a single framework consisting of scaling laws developed for the bursting of individual bubbles through two mechanisms – jet drop production for supermicron drops and submicron film drop production for the smallest drops. Because these production mechanisms are likely modulated by the organic matter present in real seawater, we now aim to apply the same framework to bursting in surfactant solutions. In this study, we consider the anionic surfactant sodium dodecyl sulfate (SDS), a well-characterized surfactant \citep{fainerman} previously used in the literature to mimic the lipids (primarily of biogenic origin) present in natural seawater (\citet{sellegri_surfactants_2006}, see discussion in \citet{burrows_oceanfilms_2014}).

The paper is organized as follows. In \S 2, we present the experimental methods and procedure for linking the bursting bubble size distribution to the emitted aerosol size distribution. In \S 3, we experimentally demonstrate the influence of surfactant on the bubble lifetime, which must be accounted for when computing aerosol production efficiency. In \S 4, we present measurements of the bubble and aerosol size distributions, and associated aerosol production efficiencies, for solutions with increasing surfactant concentration. \S 5 provides a discussion of the trends in production efficiency for different aerosol sizes.

\begin{figure}[!h]
  \centerline{\includegraphics[scale=0.77]{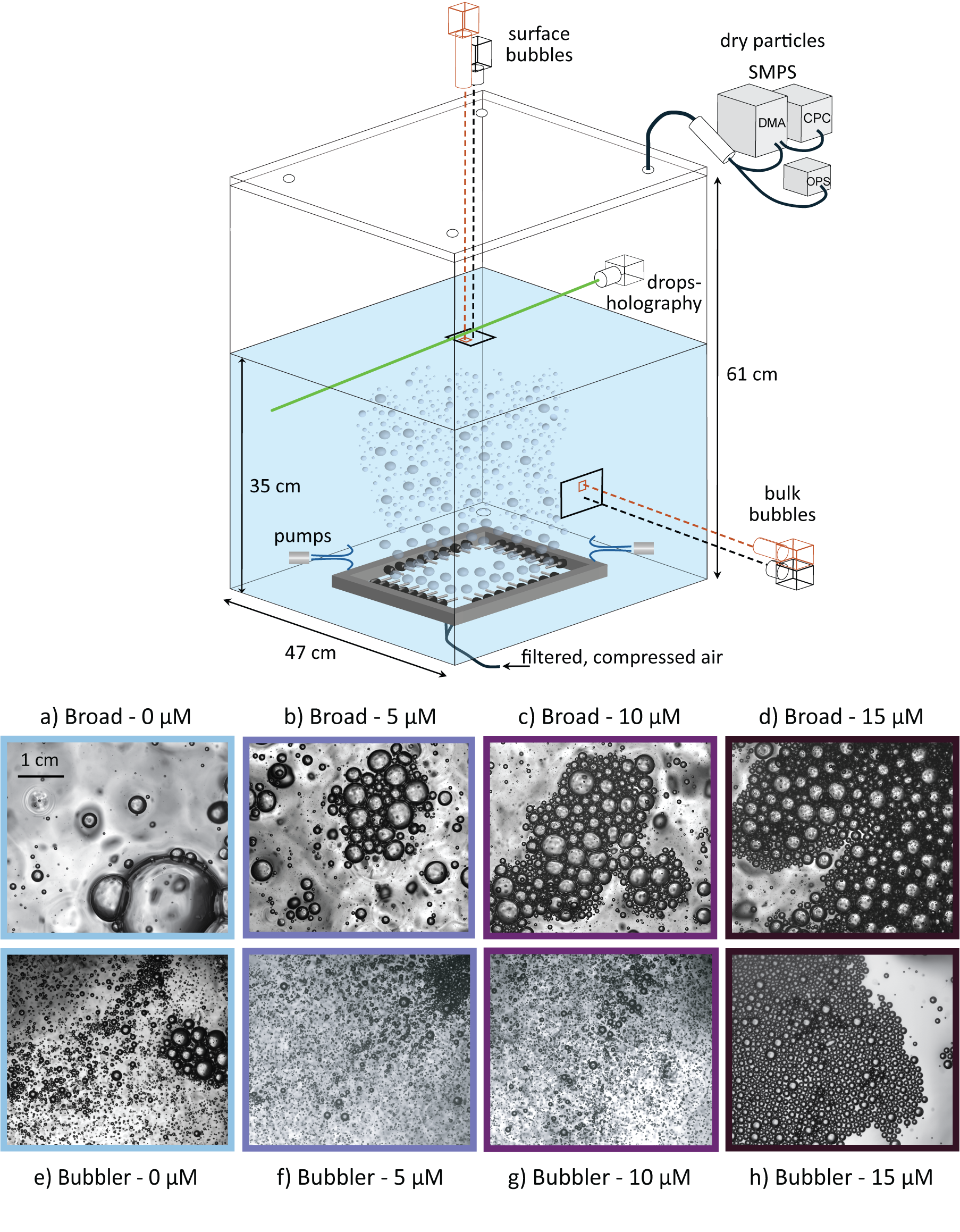}}
  \caption{Top: Sketch of bubbling tank experimental setup showing the various measurements of underwater bubbles, surface bubbles, drops, and dry particles (adapted from \citet{mazzatenta}). In the sketch, bubbles are generated as compressed air flows through a needle array; bubbles are then fragmented into many different sizes as they rise through a region of underwater turbulence (for generation corresponding to the broad-banded cases studied in the present experiments). Bottom: Representative surface images for two bubble generation methods and four surfactant concentrations. Images in the top row (a-d) correspond to the broad-banded bubble size distributions, while images in the bottom row (e-h) correspond to bubbles below 1~mm radius generated by an aquarium bubbler/porous medium. Each column corresponds to a particular concentration of sodium dodecyl sulfate (SDS) in the artificial seawater solution, ranging from 0 to 15~\textmu{M}.}
\label{fig:ims}
\end{figure}

\section{Experimental methods: Measuring bubbles, drops, and particles for solutions of various surfactant concentrations}

\subsection{Bubbling tank experimental setup}

Figure \ref{fig:ims} depicts the bubbling tank experimental setup, in which bubbles are continuously generated, rising, and bursting in a statistically-steady state. Systematic measurements of underwater bubbles, surface bubbles, drops, and dry particles are made throughout each experimental run, where the injected bubble size distribution and the solution are varied in a controlled manner. Bubbles spanning radii 30~\textmu{m} to 5~mm are measured through shadow imaging in multiple fields of view, while aerosols of dry particle diameters spanning 40~nm to 200~\textmu{m} are measured through the combination of a digital inline holographic setup (liquid drops, 10\textendash\,400~\textmu{m}), an Optical Particle Sizer (OPS, dry particles, 0.3\textendash\,10~\textmu{m}), and a Scanning Mobility Particle Sizer (SMPS, dry particles, 0.04\textendash\,0.8~\textmu{m}). The combined techniques allow us to measure both bubbles and drops/particles across a large range of sizes, which is necessary to make a quantitative link between bursting bubbles and emitted drops. The aerosol size distributions will be presented in terms of the dry particle diameter, $D_p$. Holographic measurements of liquid drop radii, $r_d$, are converted to equivalent dry particle diameters by assuming that the emitted drops contain the same concentration of salt as the bulk solution, resulting in the classic relation $D_p = 0.5\,r_d$ for artificial seawater \citep{lewis_sea_2004}. A complete description of the measurements and processing techniques are provided in \citet{mazzatenta}. Detailed bubble and aerosol measurements were made in artificial seawater solutions for five different surfactant concentrations (ranging from 0\textendash50~\textmu{M} of SDS) and two different bubble size configurations.

\subsection{Surfactant solutions and bubble configurations}

To observe the effect of surfactant on the measured bubbles and aerosols, we run the bubbling tank experiment with solutions of varying contamination. The base solution consists of deionized water mixed with ASTM D1141-98 Artificial Sea Salt (Lake Products Company LLC, Batch TL-ASTM-23101), which is composed of the same proportions of inorganic salts found in natural seawater (above 0.0004\% by weight). All solutions for this experiment were prepared with a salinity of 35~g/kg to mimic ocean water. We add sodium dodecyl sulfate (SDS) into the base solution of artificial seawater for concentrations of 5, 10, 15, and 50~\textmu{M}. In the range of surfactant concentration explored in these experiments, the static surface tension ($\gamma_0$) varies significantly, with values of $\gamma_0 = $ 71.5, 70, 65, 60, and 50~mN/m corresponding to the 0, 5, 10, 15, and 50~\textmu{M} solutions, respectively. Values of $\gamma_0$, along with full surface tension isotherms, were measured using a Langmuir trough (KSV NIMA, model KN 1003) for each case and are shown in Figure S1. Note that the presence of salts effectively reduces the critical micellar concentration \citep{salt_surfactant_2,qazi_salt}, and smaller SDS concentrations are needed to prevent coalesence due to the interaction between the anionic surfactant and the salt \citep{surfact_salt_coalesc_2,surfact_salt_stabilizations}. 

For each surfactant concentration, we test two different bubble generation configurations: (1) broad-banded bubble size distributions of rising bubbles created by fragmentation of $\sim$2~mm-radius bubbles generated through a needle array (depicted in the schematic in Figure \ref{fig:ims}) and (2) a narrower size distribution of bubbles with radii smaller than 1~mm generated using an aquarium bubbler/porous medium. The two bubble generation techniques result in significantly different surface bubble size distributions and clustering (Figure \ref{fig:ims}), which allows us to study whether the observed effects of surfactant on collective bursting are robust to surface bubble configuration. Both bubble generation techniques can be sensitive to the surfactant concentration; therefore, detailed bubble measurements are required to make a link between bursting bubbles and emitted aerosols.

Sample surface images of the two bubbling configurations with varying SDS concentrations are shown in Figure \ref{fig:ims}. As surfactant concentration is increased (from left to right in rows of images), we observe that clusters of surface bubbles become more dense as the surfactant inhibits coalescence. Note that for the broad-banded cases, 15~\textmu{M} was the highest SDS concentration that could be tested before multilayer foams began to form, and the surface became completely packed with long-lived bubbles. In the bubbler cases, a higher concentration of 50~\textmu{m} could be tested without reaching the packed state. 

\subsection{Linking bubbles and aerosols to obtain a production efficiency}

As discussed in \citet{mazzatenta}, linking the measurements of emitted drops to those of bursting bubbles at the surface or in the bulk requires information on the bubbles' journey of rising and bursting, which can be understood through a flux budget: the flux of bubbles rising through a plane underwater must be equal to the flux of bubbles living for a finite amount of time at the surface, which will control the emitted aerosols. In practice, we use the flux argument to convert the surface bubble size distribution (measured as a number of bubbles per unit area) into a quantity with the same units as the aerosol size distribution (number of aerosols per unit volume), which allows us to directly relate the bursting surface bubbles to the emitted aerosols \citep{mazzatenta}. The size distribution of bursting bubbles, $N_b(R_b)$, is then
\begin{equation}
N_b(R_b) = \frac{N_s(R_b)}{\tau_s(R_b)\ w_b(R_b)},
\label{eq:surftobulk}
\end{equation}
where $N_s$ is the measured surface bubble size distribution, and $w_b(R_b)$ and $\tau_s(R_b)$ are the size-dependent bubble rise velocity and lifetime. The resulting bursting bubble size distributions, $N_b(R_b)$, are shown for different surfactant concentrations and bubble size configurations in Figure \ref{fig:dist2}(a,c), and are used throughout the rest of the paper. 

We can verify the adequacy of our approach by confirming that the converted bursting bubble size distributions compare well with the measured bulk bubble size distributions, as shown by the comparison to the original bulk bubble measurements in Figure S2. To obtain these comparisons, we require accurate information on the bubble lifetime, which is strongly influenced by surfactant, and the rise velocity. For the rise velocity, we consider a classic parameterization for contaminated bubbles \citep{clift_bubbles_1978, takemura}. We do not expect the rise velocity to change significantly across the surfactant concentrations explored in these experiments, since the formulation already describes the rise of a contaminated bubble \citep{farsoiya_coupled_2024}.
The lifetime $\tau_s(R_b)$ for each surfactant concentration is measured by performing a raft decay experiment similar to \citet{neel2021}, described in \S 3. 

Using the converted size distribution of bursting bubbles, $N_b(R_b)$, and the aerosol size distribution, $N_d(D_p)$ (now both in units of number per unit volume), an aerosol production efficiency $E([R_{b,1}:R_{b,2}]\rightarrow [D_{p,1}:D_{p,2}])$ can then be defined to quantify the average number of aerosols ($n_d$) per number of bubbles ($n_b$), associating a specific range of bursting bubble sizes, $[R_{b,1}:R_{b,2}]$, and a corresponding range of emitted aerosol sizes, $[D_{p,1}:D_{p,2}]$: 
\begin{equation}
E([R_{b,1}:R_{b,2}]\rightarrow [D_{p,1}:D_{p,2}])=\frac{\int_{D_{p,1}}^{D_{p,2}}N_d(D_p)dD_p}{\int_{R_{b,1}}^{R_{b,2}}N_b(R_b)dR_b}\equiv \frac{n_d}{n_b}
\label{eq:eff}
\end{equation}
In the analysis in \S 4, we will focus on attributing supermicron and submicron aerosols to two specific ranges of bursting bubble sizes and investigate the effect of surfactant on each.

\section{Size-dependent bubble lifetime for various surfactant concentrations} 

\begin{figure}[!ht]
  \centerline{\includegraphics[width=0.55\linewidth]{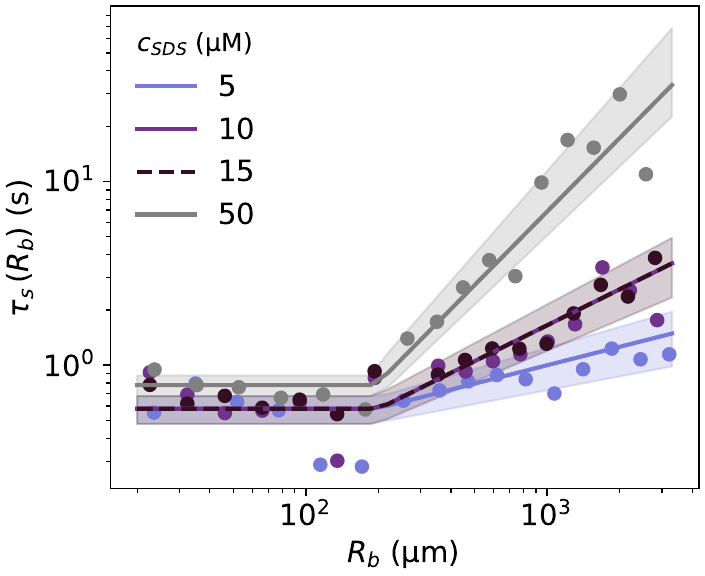}}
  \caption{Average bubble lifetime, $\tau_s\,(R_b)$, as a function of the bubble radius, $R_b$, for solutions of various surfactant concentration, $c_{\textit {\tiny SDS}}$. Circular markers show the data from the small-scale raft decay experiments, with each point representing the average of $\sim$500 bubbles. The solid line shows the size-dependent lifetime relation that we use to construct the bursting bubble size distribution shown in Figure \ref{fig:dist2} (for the main bubbling tank experiment), and the shaded regions indicate approximate uncertainties on the bubble lifetime. Note that the same lifetime trend is taken for both the 5 and 10~\textmu{M} cases.} 
\label{fig:lifetime}
\end{figure}
In order to disentangle the effects of surfactant on aerosol generation, we first need to account for surfactant effects on other surface bubble processes, notably on the surface bubble lifetime (the average time that a bubble resides at the surface before bursting).  
To provide a cohesive description of the dependence of bubble lifetime on bubble size and surfactant contamination in a polydisperse raft, we revisit the raft decay experiments of \citet{neel2021}, now generating polydisperse bubble rafts and measuring bubbles down to the smallest sizes ($R_b\sim$ 20~\textmu{m}). 
In the small-scale raft decay experiment, repeatable rafts are formed by impinging jets, bubbles are tracked using two cameras with overlapping fields of view, and ensemble-averaged lifetimes are extracted for bubbles in various size bins.
Resulting data for the average bubble lifetime, $\tau_s (R_b)$, is shown as a function of bubble radius in Figure \ref{fig:lifetime} (circle markers), showing an overall dependence on both SDS concentration and bubble size. 

For bubbles with radii $R_b <$ 200~\textmu{m}, changes in the bubble lifetime with both bubble size and surfactant remain small. For bubbles of radii $R_b >$ 200~\textmu{m}, bubble lifetime becomes dependent on both surfactant concentration and bubble size; in this regime, larger bubbles are found to have longer lifetimes than smaller bubbles, and the trend steepens with increasing surfactant concentration. The lines show the lifetime relations used to convert the surface bubble size distribution (following Equation \ref{eq:surftobulk}) for each surfactant concentration. For the case of artificial seawater with no surfactant, we take a constant lifetime of $\tau_s$ = 0.4~s, coherent with the bulk bubble comparison and previous observations \citep{mazzatenta,tristan_bubble}.
These size-dependent lifetime results have important practical implications when calculating the actual efficiency of aerosol production, as an increased lifetime due to surfactant would lead to an artificially-inflated efficiency if not accounted for (effectively combining Equations \ref{eq:surftobulk} and \ref{eq:eff}).

\section{Surfactant effect on aerosol production efficiencies} 

\begin{figure}[!ht]
  \centerline{\includegraphics[width=0.93\linewidth]{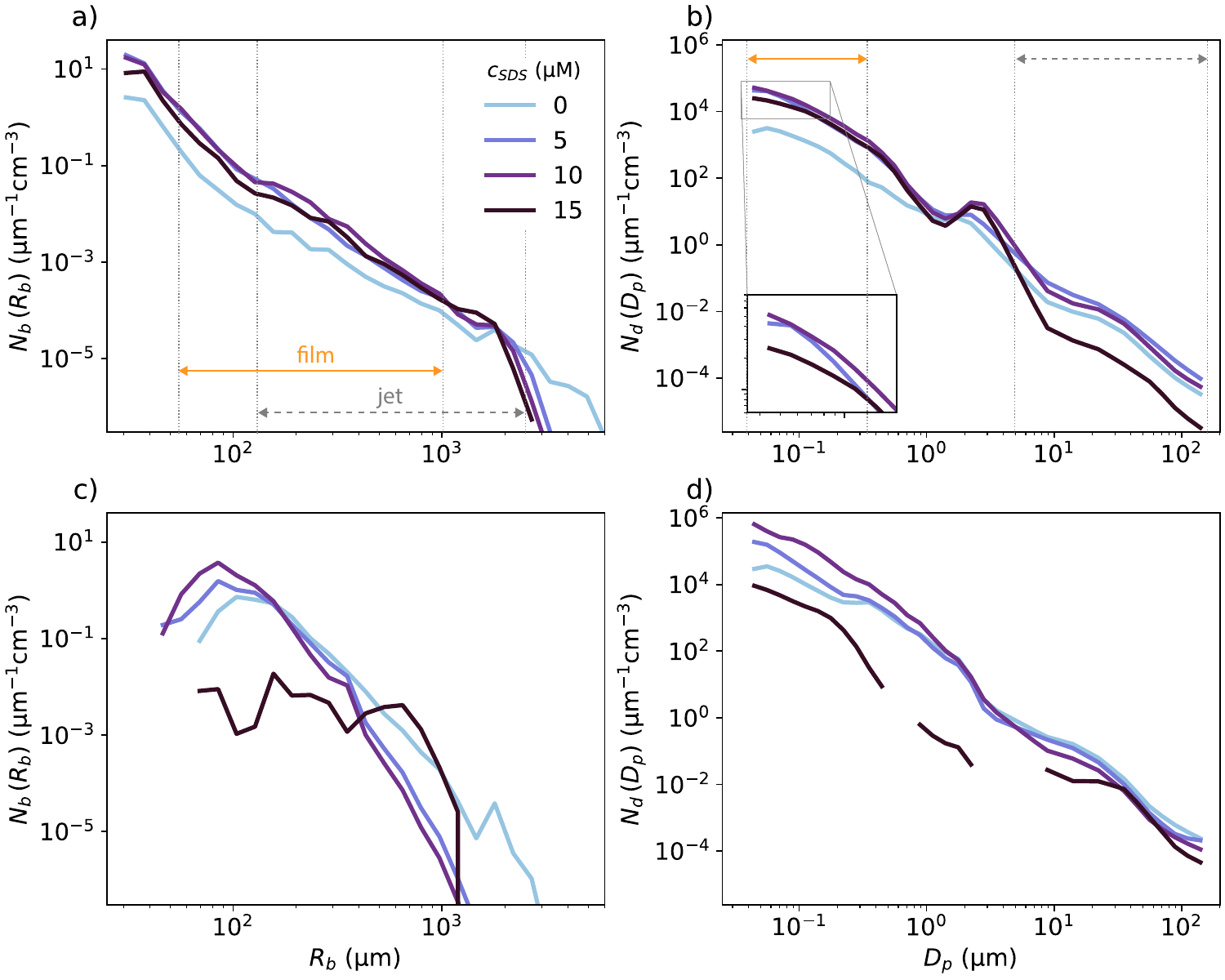}}
  \caption{Bubble size distributions ($N_b(R_b)$, (a,c)) and aerosol size distributions ($N_d(D_p)$, (b, d), in terms of the dry particle diameter, $D_p$) for the broad-banded (a,b) and bubbler (c,d) generation cases for increasing surfactant concentration ($c_{\textit {\tiny SDS}}$, colors). The vertical dashed lines indicate the size ranges in which the distributions are integrated to calculate aerosol production efficiencies for submicron film drop (orange) and jet drop (gray) production.}
\label{fig:dist2}
\end{figure}

We first discuss the trends in the bubble and aerosol size distributions, shown in Figure \ref{fig:dist2} for the broad-banded (a,b) and bubbler (c,d) cases. For the bubbles of the broad-banded cases (panel a), we see an increase in the number of small bubbles when surfactant is added. These small bubbles are formed during underwater bubble breakup in turbulence and then rise to the surface (see also bubble breakup experiments from \citet{zhan_tristan}). The bubbler (porous media) generation underwater is also sensitive to surfactant; we observe an increase in small bubbles for the 5 and 10~\textmu{M} cases compared to the case with no surfactant (panel c), and then very few small bubbles for the 15~\textmu{M} case (which has a significant population of bubbles around $R_b$ = 750~\textmu{m}). 
In both bubble configurations, we observe a decrease in the number of very large bubbles once surfactant is added, as the surfactant inhibits coalescence. 

The aerosol size distributions (b,d) are shown in terms of the dry particle diameter, $D_p$. Looking qualitatively at the distributions, we observe an increase in number of submicron drops as the SDS concentration is increased up to 10~\textmu{M}, for both the broad-banded and bubbler cases. As the concentration is increased further to 15~\textmu{M}, the submicron portion of the distribution falls, with a more significant decrease in the bubbler case. For larger aerosols ($D_p >$ 5~\textmu{m}), we observe a significant decrease in the aerosol concentration for the 15~\textmu{M} concentration, especially in the broad-banded case. 

\begin{figure}[!h]
  \centerline{\includegraphics[width=0.96\linewidth]{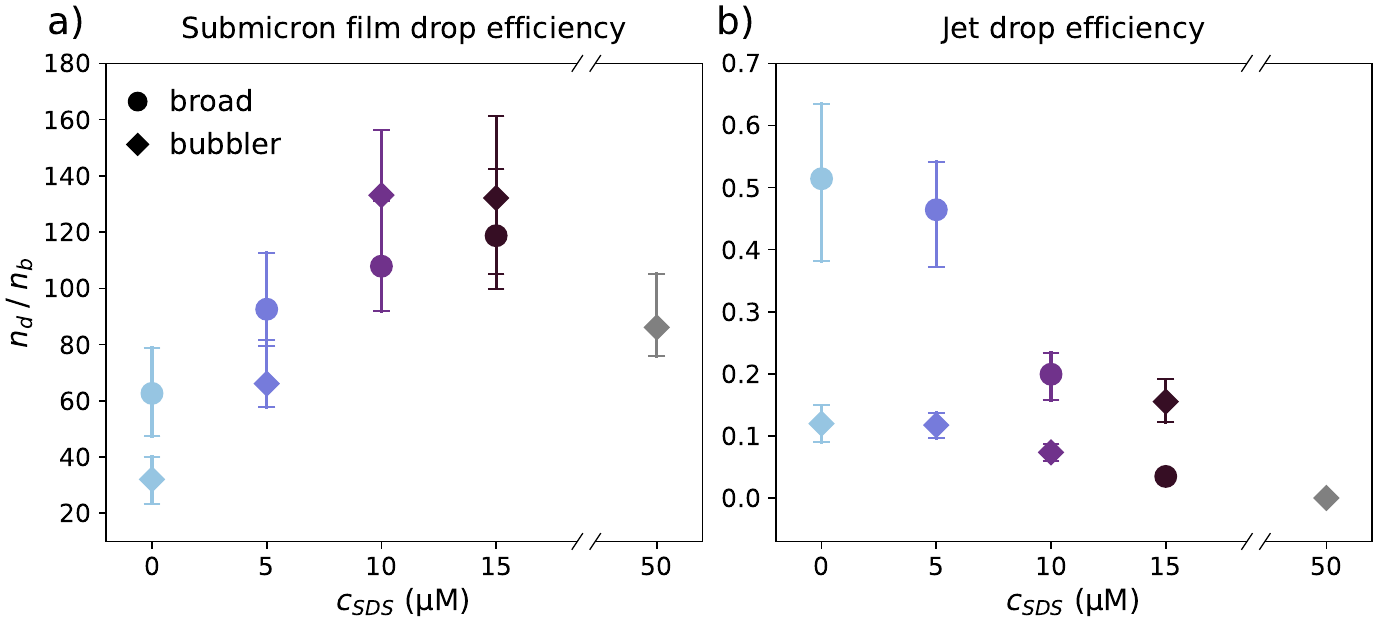}}
  \caption{Aerosol production efficiencies ($n_d/n_b$, number of aerosols per bubble) with increasing SDS concentration ($c_{\textit {\tiny SDS}}$) for bubble/aerosol size ranges corresponding to primarily (a) film drop production and (b) jet drop production for both the broad-banded (circles) and bubbler (diamonds) cases. (a): Aerosol production efficiency for bursting bubbles in the range $R_b$ = [55, 1000]~\textmu{m}, which could produce film drops of dry particle diameter, $D_p<$ 0.5~\textmu{m}. (b): Production efficiencies for bubbles in the range $R_b$ = [130, 2500]~\textmu{m}, which produce jet drops of $D_p >$ 5~\textmu{m}. Error bars represent the estimated uncertainties on the efficiency that stem from the choice of size-dependent bubble lifetime, where the upper and lower lifetime bounds shaded in Figure \ref{fig:lifetime} are used for $\tau_s(R_b)$ in Equation \ref{eq:surftobulk} to convert the surface bubble size distribution. No data was collected for the broad-banded case (circles) at the 50~\textmu{M} concentration, as multiple layers of bubbles began to stack on the surface in this configuration.} 
\label{fig:eff}
\end{figure}

With measurements of both bubbles and aerosols down to the smallest sizes, we can make a quantitative link between the bursting bubbles and emitted aerosols. To calculate the aerosol production efficiency, $n_d/n_b$, we integrate both the aerosol size distribution and the bubble size distribution in particular size ranges (following Equation \ref{eq:eff}, with size ranges illustrated in Figure \ref{fig:dist2}) for each case. The choices of bubble/aerosol size ranges we consider are based on understanding in the literature for bursting of an individual bubble of a given size through jet drop and submicron film drop production \citet{mazzatenta, deike_mechanistic_2022}.  
Following the scaling laws for jet drop production developed based on data from experiments and simulations, we consider that particles of diameter $D_p >$ 5~\textmu{m} are produced by bubbles in the radius range $R_b$ = [130, 2500]~\textmu{m} \citep{ganan-calvo_revision_2017,blancorodriguez_sea_2020,spiel_jet_97,ghabache_physics_2014,ghabache_size_2016,walls_jet_2015,brasz_minimum_2018,berny2021}. Looking at jet drop production efficiencies for the case with no surfactant (0~\textmu{M}, light blue circle and diamond in Figure \ref{fig:eff}b), we find values around 0.6 and 0.1 drops per bubble for the broad-banded and bubbler cases, respectively; coherent with our past work using the same bubbling tank setup \citep{mazzatenta}, and similar efficiencies around 0.5 drops per bubble in collective bursting experiments from \citet{wang2017, wang_jet}. 
We then consider that submicron aerosols with $D_p <$ 0.5~\textmu{m} are produced by bubbles in the range $R_b$ = [55, 1000]~\textmu{m} through a film drop production mechanism. For the case without surfactant, we find efficiencies between 35-65 drops per bubble (Figure 4a), which is in line with data reported in the literature for submicron film drop production \citep{jiang2022, jiang_abyss_2024, chu25,mazzatenta}). 

Figure 4 demonstrates that surfactant contamination has different effects on the production efficiency of submicron compared to supermicron aerosols, which is already evident from Figure \ref{fig:dist2}. To extract trends with increasing surfactant concentration, we focus on these smallest ($D_p <$ 0.5~\textmu{m}) and largest  ($D_p >$ 5~\textmu{m}) aerosol size ranges. We consider the small size range to be primarily film drops and the large size range to be primarily jet drops, although there may be some overlap in the mechanism by which these drops are produced. The analysis of aerosol production efficiencies in two size ranges is supported by the full attribution from the individual bursting scaling laws, which is detailed in \citet{mazzatenta} and Text S3, and demonstrated in Figures S3 and S4 of the Supplementary Information. 

Looking first at the production of submicron aerosols ($D_p <$ 0.5~\textmu{m}) by bubbles of radii between 55~\textmu{m} and 1~mm (Figure \ref{fig:eff}a), we observe an increase in the aerosol production efficiency as SDS concentration increases up to 10 or 15~\textmu{M} for the bubbler (diamonds) and broad-banded (circles) cases, respectively. At these concentrations of 10 to 15~\textmu{M}, the efficiency reaches values between 120 and 150 aerosols per bubble, corresponding to an increase of a factor 2 to 4 when compared to the cases without surfactant. An additional condition at 50~\textmu{M} was run for the bubbler configuration; we observe that the submicron aerosol production efficiency drops off for this highest surfactant concentration, showing that there exists an optimal surfactant concentration for which production efficiency is the highest.
Next we consider the production of large jet drops ($D_p >$ 5~\textmu{m}) by bursting bubbles of radii between 130~\textmu{m} and 2.5~mm (Figure \ref{fig:eff}b). For the broad-banded case, the aerosol production efficiency decreases significantly with increased surfactant concentration \textendash\, the opposite of the trend observed for submicron aerosol production. The bubbler cases show very little trend in jet drop production efficiency in this size range as SDS concentration is increased through 15~\textmu{M}. For the 50~\textmu{M} bubbler case, production of large drops is shut down entirely, and no aerosols larger than 5~\textmu{m} are generated. 

While the exact magnitudes in efficiencies are sensitive to the selected ranges of bubble and aerosol sizes and estimated lifetime, the trends in efficiency for different aerosol sizes are robust: as SDS concentration increases, the production efficiency increases for submicron aerosols up to a certain concentration, and decreases for supermicron aerosols. Surfactant contamination therefore affects aerosol production in a way that cannot be explained through changes in the bubble size distribution. The different trends observed with surfactant between the film drop and jet drop production ranges indicate that SDS likely affects the two drop production mechanisms in different ways.

\section{Discussion and conclusion}

We have experimentally studied the effect of SDS surfactant on collective bubble bursting, probing various regimes of surface bubble behavior using five SDS concentrations and two bubble size configurations. Through measurements of bubbles and aerosols down to small sizes, we found aerosol production efficiencies and observed that: 
1) For surfactant solutions mixed with sea salt, the surface bubble lifetime is dependent on both surfactant concentration and bubble size, which must be accounted for to accurately compute aerosol production efficiencies.
2) SDS influences aerosol production differently for aerosols of different sizes: production efficiency of submicron aerosols ($D_p <$ 0.5~\textmu{m}) increases with surfactant up to a factor 2 to 4 to reach an optimal efficiency at a certain concentration (likely determined by the specific surfactant), while production of the large aerosols ($D_p >$ 5~\textmu{m}) decreases with contamination. We also note that in the bubbler cases (i.e. only small bubbles), we see an increase in submicron aerosol production efficiency with surfactant concentration even if the surface clustering is not modified, implying that surfactant affects the bursting process outside of its influence over coalescence and surface packing.

To think about the different trends in production efficiency for different aerosol size ranges, we can consider how surfactant may affect the jet drop versus film drop production. For jet drop production (which dominates the supermicron drop production), the addition of surfactant results in surface tension gradients along the bubble cap, which induce Marangoni stresses that can lead to the emission of either smaller drops \citep{constante,pierre,pico,vega} or potentially fewer drops \citep{jun_tristan}, depending on the bubble size (both of which might explain our experimental results). For the largest concentrations considered, when production is effectively suppressed, packing and foam inhibit jet formation. Expected effects of surfactant on the mechanism of film drop production are less clear, but the increase in lifetime with surfactant may cause bubbles to burst with a thinner film thickness, which may destabilize into numerous small drops.

We have found that high-quality bubble measurements, in addition to aerosol measurements, are needed to make quantitative statements regarding aerosol production efficiency as surfactant contamination is increased. With these measurements, we can begin to disentangle the effect of surfactant on the bursting process, beyond its effects on bubble generation and surface bubble clustering. Future work could involve experiments with different surfactants relevant to the ocean context and should connect to the actual composition of the ejected aerosol (following ideas in \citet{prather2013,wang2017}). Such data would then inform how modifications to the bursting process affect the size and number but also composition of emitted aerosols, for realistic contaminated seawater solutions, eventually leading to improved sea spray emission functions accounting for the seawater composition.

\vspace{10pt} 

\noindent\textbf{Funding:}
This work was supported by the National Science Foundation under grant number 2318816 to LD; the Princeton Catalysis Initiative; the National Science Foundation Graduate Research Fellowship to MM; and the High Meadows Environmental Institute Hack Award to MM.

\vspace{5pt} 
\noindent The authors report no conflict of interest.

\vspace{5pt}
\noindent\textbf{Data availability statement:}
The experimental data used to prepare this work is publicly available through the Princeton Data Commons via \url{https://doi.org/10.34770/fjb4-ts52}.

\vspace{5pt} 

\noindent\textbf{Author ORCIDs:}
M. Mazzatenta, https://orcid.org/0000-0002-8362-0349; S.M. Koblensky, https://orcid.org/0009-0007-8535-9571; L. Deike, https://orcid.org/0000-0002-4644-9909.


\bibliographystyle{biblio}
\bibliography{biblio}

\newpage

\section*{\centering Supplementary Information for ``Surfactant effect on collective bubble bursting and aerosol emission"}

\section*{S1 Representing Aerosol Size Distributions in Terms of Dry Particle Diameter}

The aerosol size distributions are presented in Figure 3 of the manuscript in terms of the dry particle diameter, $D_p$. Liquid drops measured by the holographic setup are converted to an equivalent dry particle diameter by assuming that the drop and bulk solution contain the same concentration of salt, following the relation $D_p = 2\,r_d\,[(\rho_{salt}\,/\,\rho_{water})\,(1000\,/\,S)]^{-1/3}$ (where $r_d$ is the liquid drop radius, $D_p$ is the dry particle diameter, and $S$ is the salinity of the solution in g/kg). For the artificial seawater solutions used in this experiment (35~g/kg, $\rho_{salt} =$ 2,056~g/L), the conversion $D_p = 0.5\,r_d$ is used \citep{lewis_sea_2004}. Details of the overlapping SMPS, OPS, and holographic measurements used to construct these distributions are provided in \citet{mazzatenta}. 

\section*{S2 SDS Concentrations Compared to Previous Studies}

The concentrations of SDS used in this experiment are low compared to previous studies involving solutions of SDS with no salt \citep{neel_surf_2022}, but they are comparable to concentrations used in bubble raft experiments involving solutions of SDS with sea salt \citep{sellegri_surfactants_2006}, which is in line with the idea that the presence of salts effectively reduces the critical micellar concentration \citep{salt_surfactant_2,qazi_salt}, and that smaller SDS concentrations are needed to prevent coalesence due to the interaction between the anionic surfactant and the salt \citep{surfact_salt_coalesc_2,surfact_salt_stabilizations}. Values of $\gamma_0$, along with full surface tension isotherms, were measured using a Langmuir trough (KSV NIMA, model KN 1003) for each case and are shown in Figure S1. Note that for the artificial seawater case with no surfactant, the static surface tension of 71.5~mN/m is very close to that of clean water (72~mN/m), confirming that the base solution is composed mostly of inorganic salt mixture.  

\section*{S3 Individual Bubble Bursting Scaling Laws}

In the manuscript, aerosol production efficiencies are calculated in different bubble/drop size ranges \textendash\, one corresponding to submicron film drop production, and one corresponding to jet drop production. To determine these size ranges, we choose aerosol size ranges of interest and then follow scaling laws developed for the bursting of individual bubbles (not collective) to choose the associated range of bursting bubble sizes. Table S1 shows the equations based on scaling laws developed for individual bubble bursting, which are used to link the measured bubbles and aerosols. The formulations provided for the mean radius of drops, $\langle r_{d} \rangle(R_b)$, emitted by bubbles of a certain size are used to determine the film drop and jet drop production size ranges labeled on the bubble/drop size distributions in Figure 3 of the manuscript. Note that while the full range of bubble sizes capable of producing jet drops is given in Table S1, Figure 3 of the manuscript highlights a smaller range of bubbles sizes ($R_b =$ [130,2500]~\textmu{m}), as this is the range that corresponds to the drop size range of interest in this study. 

The full attribution, linking emitted drops to bursting bubbles through film drop and jet drop production mechanisms, is shown in Figures S3 and S4 (plotted in terms of the liquid drop radius, $r_d$). To predict the drops produced by the ensemble of bursting bubbles, we consider that the drop size distribution $N_d(r_d)$ is obtained by integrating individual bursting scaling laws over the size distribution of bursting bubbles $N_b(R_b)$ in the size range [$R_{b1}$, $R_{b2}$]. This amounts to the formulation:
\begin{equation}
  N_d(r_d) = \int_{R_{b1}}^{R_{b2}} \frac{N_b(R_b)n_{d}(R_b)}{\langle r_d \rangle(R_b)} p\left(\frac{r_d}{\langle r_d \rangle}, R_b\right) \,dR_b\, ,
  \label{globaldrop}
\end{equation}
an approach initially proposed by \citet{lv2012} for production of film drops through a centrifuge mechanism and later extended to jet drops \citep{berny2021} and submicron film drops \citep{deike_mechanistic_2022}. Here $N_b(R_b)$ is the measured bubble bursting size distribution, $\langle r_d \rangle(R_b)$ is the mean drop size produced by a bursting bubble of size $R_b$, $n_{d}(R_b)$ is the number of drop produced by a bubble bursting of size $R_b$, and $p(r_d/\langle r_d \rangle, R_b)$ is the probability density function of drop sizes produced by a bubble of size $R_b$, assumed to be a Gamma distribution of known order \citep{lv2012,berny2021,deike_mechanistic_2022}.

We use all of the formulations provided in Table S1 to compute the drop size distributions shown in Table S1, compared to the measured drop size distributions. The individual bubble scalings and details of the attribution are described fully in \citet{mazzatenta}. Parameters for submicron film drop production differ slightly from those provided in \citet{mazzatenta}, adjusted slightly so that the case with no surfactant is described well by the scaling laws. Full details and discussion of the sensitivity of the predicted drop size distributions to different choices of $R_{b1}$, $R_{b2}$, and the order of the Gamma distribution are discussed in \citet{mazzatenta}.

\begin{figure}[!h]
  \centerline{\includegraphics[scale=1]{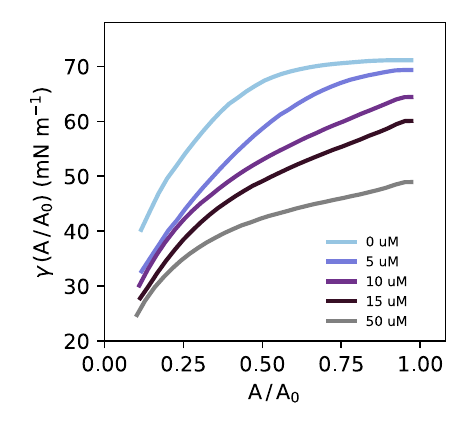}}
  \caption*{Figure S1: Surface tension isotherms measured using the Langmuir trough for solutions of artificial seawater and all SDS concentrations presented in the manuscript. The surface tension is plotted as a function of the portion of the trough surface area, $A_0$, that remains uncompressed as the trough barriers move to compress the solution. The value located at $A/A_0 =$ 1.00 corresponds to the static surface tension, $\gamma_0$, for each case.} 
\label{fig:langmuir2}
\end{figure}

\begin{figure}
  \centerline{\includegraphics[scale=0.65]{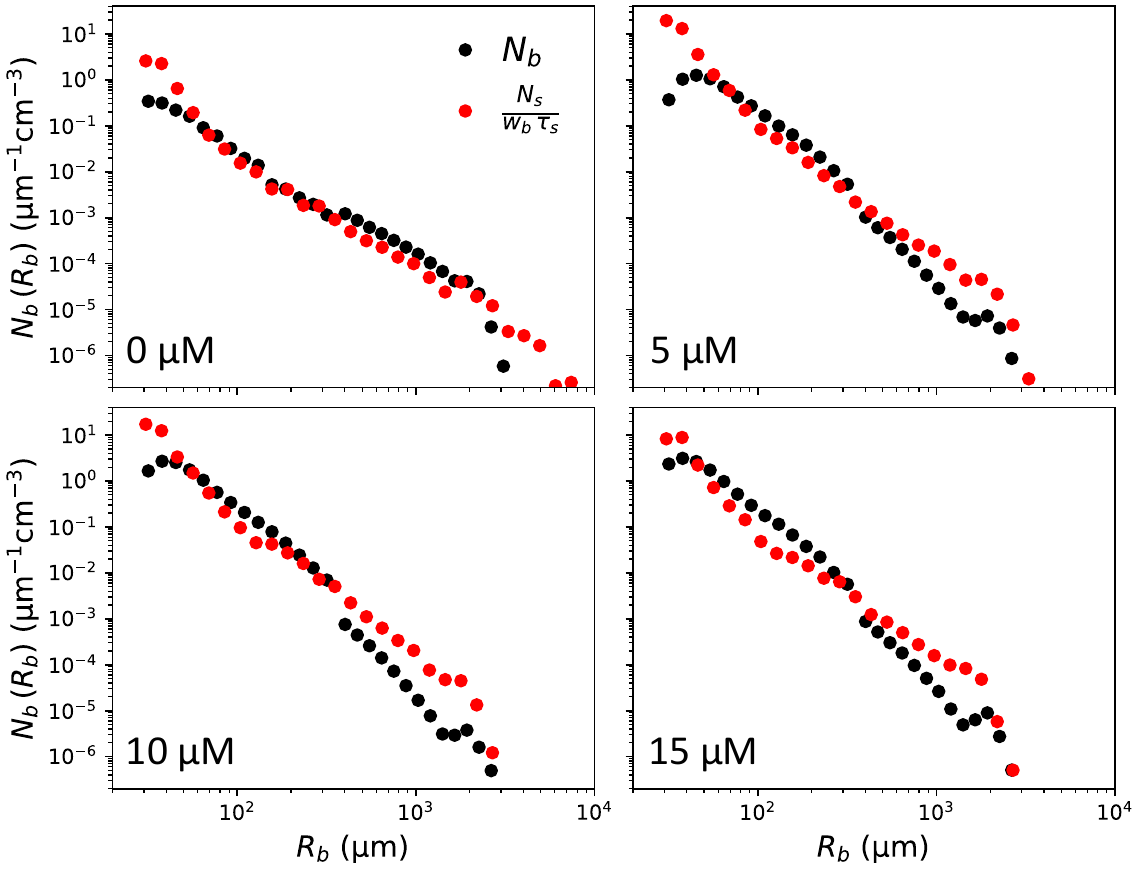}}
  \caption*{Figure S2: Comparison between the bulk bubble size distribution, $N_b$, (measured directly underwater), and the converted surface bubble size distribution, $N_s(R_b)\,/\,[\,w_b(R_b)\, \tau_s(R_b) \,]$, where $N_s$ is the size distribution measured directly at the surface, and $w_b(R_b)$ and $\tau_s(R_b)$ are the size-dependent bubble rise velocity and lifetime, respectively. The bubble lifetime varies with the surfactant concentration, as discussed in \S 2.3 of the manuscript. \S 2.3 also described the flux argument used to convert the surface measurement. The comparisons are shown here for the broad-banded bubble size configuration for SDS concentrations ranging 0 to 15~\textmu{M}.}
\label{fig:flux2}
\end{figure}

\begin{table}
  \begin{center}
\def~{\hphantom{0}}
  \begin{tabular}{l|c@{}cc@{}@{}c}
      Mechanism  & [$R_{b1}$, $R_{b2}]$ (\textmu{m}) & $\langle r_{d} \rangle(R_b)$ (\textmu{m}) & $n_d(R_b)$ & Order of $\Gamma$ \\[1pt]
        &  &  & & distribution \\[1ex]
        \hline
        & & & &  \\[-1ex]
       Film (submicron) & [55, 1000] & $0.40\ \left(\frac{R_b}{l_c}\right)^{1/3}$ & 60 & 6\\ [2pt] 
       Jet & [30, 2500] & $0.60\ l_{\mu}\ \left( \sqrt{La}\left(\sqrt{\frac{La}{La_*}} - 1\right)\right)^{5/4}$ &
  $30 \left(\frac{R_b}{l_{\mu}}\right)^{-1/3}$ & 11\\[2pt] 
       \\[2pt]
  \end{tabular}
  \caption*{Table S1: Adapted from \citet{mazzatenta}. Equations and parameters used to calculate the drop size distribution $N_d(r_d)$ from input/measured bubble size distribution $N_b(R_b)$, following (with adaptation) the scaling laws developed for individual bubble bursting through submicron film drop and jet drop production mechanisms. Equations are shown for the mean radius $\langle r_{d} \rangle(R_b)$ and number $n_d(R_b)$ of drops emitted by a bubble of radius $R_b$, as well as the radius bounds [$R_{b1}$, $R_{b2}$] and the order of the Gamma distribution (assumed to represent the distribution of drop sizes $p(r_d/\langle r_d \rangle, R_b)$). The parameters above are required to compute $N_d(r_d)$ for an ensemble of bursting bubbles, following the equation provided in Text S1 above. For submicron film drop production, the size scaling is based on the data from \citet{jiang2022} (also used in \citet{chu25} and discussed in \citet{deike_mechanistic_2022}). The drop number is taken as a single value of 60 drops/bubble, based on \citet{jiang2022}. For jet drop production, we use the formulation for the first drop size given by \citet{ganan-calvo_revision_2017}, and the drop number scaling from \citet{berny2021} is used with a modified prefactor suggested by \citet{wang2017, mazzatenta}. The Laplace number $La = R_b/l_{\mu}$ controls jet drop ejection, comparing the radius of the bursting bubble, $R_b$, to the visco-capillary length, $l_{\mu} = \mu^2/\/\rho\gamma$ ($\mu$ is water viscosity, $\rho$ is water density). $La_* = 550$ is taken for the critical Laplace number for jet drop formation \citep{walls_jet_2015, berny_role_2020}. Full details of the parameters above are discussed in \citet{mazzatenta}.}
  \label{tab:paramsfull}
  \end{center}
\end{table}

\begin{figure}
  \centerline{\includegraphics[scale=0.55]{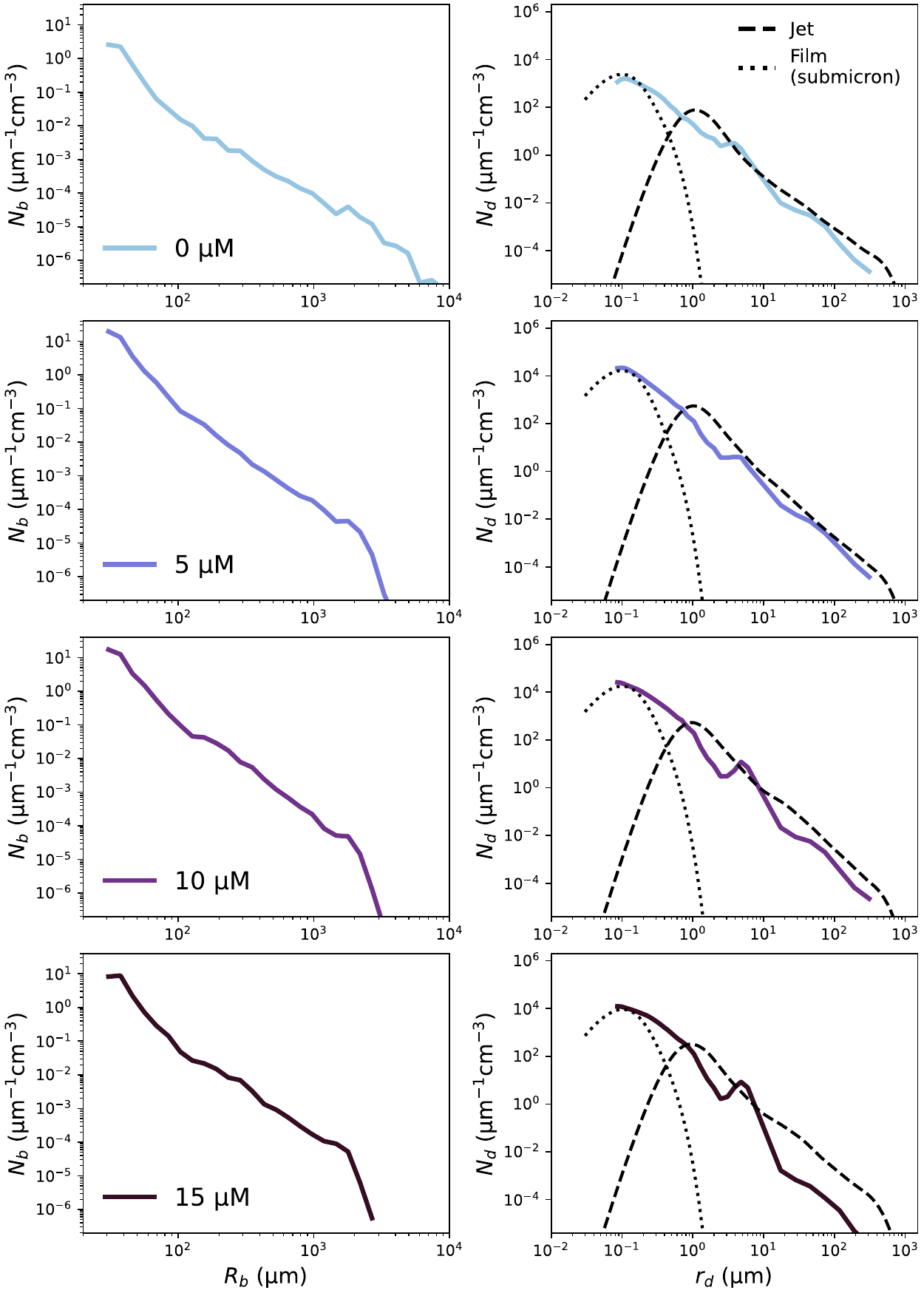}}
  \caption*{Figure S3: Measured bursting bubble size distributions ($N_b(R_b)$) and drop size distributions ($N_d(r_d)$, plotted in terms of the liquid drop radius) (solid, colored line) for the broad-banded cases at the different surfactant concentrations. The predicted drop size distributions from individual bubble bursting scaling laws are plotted in different styles for jet (dashed) and submicron film (dotted) drop production. See Text S1 and \citet{mazzatenta} for more information on computing the predicted drop size distributions using individual bubble bursting scaling laws.}
\label{fig:fwdbroad}
\end{figure}

\begin{figure}
  \centerline{\includegraphics[scale=0.53]{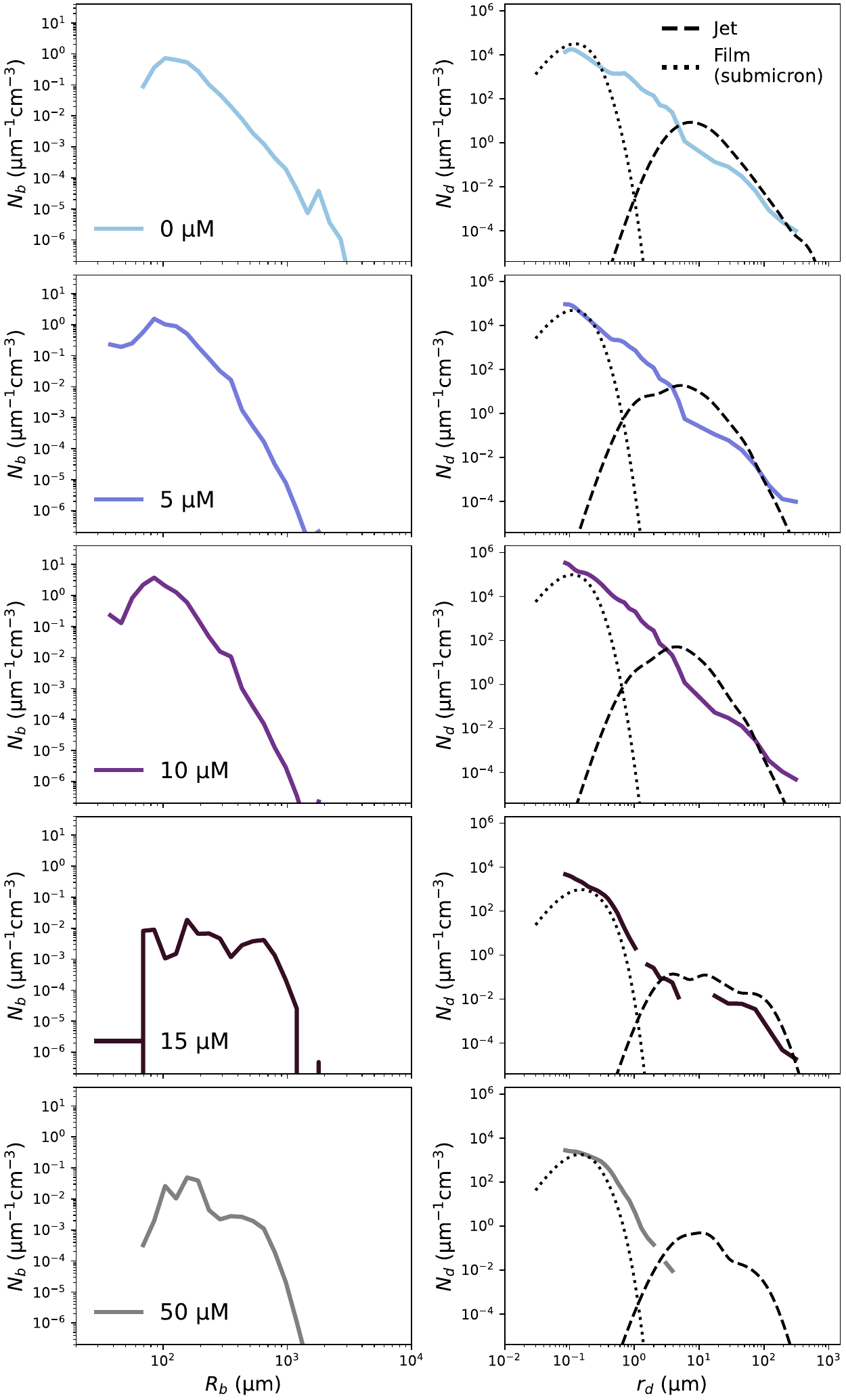}}
  \caption*{Figure S4: Measured bursting bubble ($N_b(R_b)$) and liquid drop ($N_d(r_d)$) size distributions (solid, colored line) for the bubbler cases at the different surfactant concentrations. The predicted drop size distributions from individual bubble bursting scaling laws are plotted in different styles for jet (dashed) and submicron film (dotted) drop production. See Text S1 and \citet{mazzatenta} for more information on computing the predicted drop size distributions using individual bubble bursting scaling laws.}
\label{fig:fwdbubbler}
\end{figure}

\end{document}